\documentclass[preprint]{aastex}
\usepackage{epsfig}

\begin{document}

\title{Cygnus A}

\author{A. S. Wilson, K. A. Arnaud\altaffilmark{1}, D. A. 
Smith, Y. Terashima\altaffilmark{2}, A. J. Young}

\affil {Astronomy Department, University of Maryland, College Park, MD 20742,
U.S.A.}
\altaffiltext{1}{Laboratory for High Energy Astrophysics, NASA/GSFC, 
Code 662, Greenbelt,
MD 20771, U.S.A.}
\altaffiltext{2}{Institute of Space and Aeronautical Science, 3-1-1 Yoshinodai,
Sagamihara, Kanagawa 229-8510, Japan}

\begin{abstract}
We report Chandra imaging-spectroscopy and RXTE spectroscopy of the nearby, 
powerful radio galaxy
Cygnus A. Various aspects of the results are discussed, including the
X-ray properties
of the nucleus, the radio hot spots, the cluster of galaxies, the
prolate cavity in
the ICM inflated by the radio jets and ``bands'' of thermal gas which encircle
the cavity in its equatorial plane. The hard X-ray emission of the nucleus
extends to 100 keV and originates from an unresolved source absorbed by a large
column density
(N$_{\rm H}$ $\simeq$ 2 $\times$ 10$^{23}$ cm$^{-2}$) of gas.
The soft ($<$ 2 keV)
nuclear emission exhibits a bi-polar structure which extends $\simeq$ 2 kpc 
from
the nucleus and is strongly correlated with both optical continuum and
emission-line morphologies. It is suggested that this nebulosity is
photoionized by the nucleus and that the extended X-rays are electron-scattered
nuclear radiation. All four radio hot spots are detected in X-rays, with
the emission resulting from synchrotron self-Compton radiation in an
approximately equipartition field. The temperature of the X-ray emitting
intracluster gas drops from $\simeq 8$~keV more than $100$~kpc from
the center to $\simeq 5$~keV some $80$~kpc from the center, with the
coolest gas immediately adjacent to the radio galaxy. 
There is a metallicity gradient in the X-ray emitting gas,
with the highest metallicities ($\sim$ solar) found close to the
center, decreasing to $\sim 0.3$~solar in the outer parts. We have
used the assumption of hydrostatic equilibrium to derive a total
cluster mass within 500~kpc of $2.0 \times 10^{14} \, M_{\odot}$ and
$2.8 \times 10^{14} \, M_{\odot}$, and a gas fraction in the cluster
within $500$~kpc of
$0.055$ and $0.039$ for a constant and centrally decreasing temperature
profile, respectively. We show that the limb-brightened edge of the cavity is
hotter than the nearby, innermost region of the cluster gas, indicating heating
by the
expanding, jet-driven cavity. Conversely, the ``bands'', interpreted as gas 
being accreted onto the galaxy, are cooler. Within the cavity, 
there is evidence for diffuse X-ray
emission, in addition to the X-ray emissions
related to the jets and hot spots. 

\end{abstract}

\keywords{galaxies: active -- galaxies: clusters -- galaxies: individual 
(Cygnus A) -- galaxies: nuclei -- galaxies: X-rays --
intergalactic medium -- interstellar medium}

\section{Introduction}
  
Cygnus A is a powerful radio galaxy that is unusually close to us (z = 0.0562,
Stockton et al. 1994). Its radio luminosity is more than two orders of
magnitude above the Fanaroff-Riley I/II boundary and 1.5 orders of
magnitude more luminous than any other source at z $\le$ 0.1 (Carilli \&
Barthel 1996). Most sources with similar radio luminosities are found at 
z = 1 and beyond. For these reasons, Cyg A is considered the archetypal
powerful radio galaxy and has been extensively studied in all wavebands.

In this paper, we summarise results from a study of Cyg A with the Chandra
X-ray Observatory. Cyg A was observed with the spectroscopic array of CCDs,
with the nucleus at the aim point on chip S3;
the date of observation was chosen so that the extension some 15\arcmin\ to the
NW (Arnaud et al. 1984; Reynolds \& Fabian 1996) was also imaged. An image of 
the 6
CCDs which were read out is shown in Fig. 1. The Cyg A cluster is the bright
region near the center and the putative other cluster is seen to the NW.

An enlargement of the inner region, on the scale of the double radio source,
is given in Fig. 2 and shows a wealth of structure. The brightest region, in the
center, is the galaxy nucleus. The ``hot spots'' (two bright ones about 
1\arcmin\ to the NW and one bright and one faint one a similar distance
to the SE) coincide with the radio hot spots. There are two diffuse, linear 
structures running from the nucleus towards each pair of hot spots;
these quasi-linear features seem to
be associated with the radio jets, though in X-rays the emission is more 
diffuse. The overall, prolate shape presumably represents the
limb-brightened cavity inflated in the intracluster medium (ICM) by 
relativistic particles that have
escaped from the hot spots, as predicted by Scheuer (1974). 
The whole structure is
embedded in the ICM of the cluster of galaxies, which extends to a much larger
scale (see Fig. 1).
Lastly, curved ``bands''
run transversely (i.e. NE - SW) across the cavity and appear to be the 
projections
of rings of gas which encircle it; if the plane of these rings is perpendicular
to
the radio axis, the rings must be predominantly on the far side of the cavity
since the jet to the NW is the nearer. We discuss these various features in
turn and refer the reader to Wilson et al. (2000, Paper I, about the ``hot 
spots''), Young et al. (2002,
Paper II, about the nucleus) and Smith et al. (2002, Paper III, about the 
cluster) for further details.

\section{The nucleus}

Fig. 3 shows a color representation of the Chandra X-ray image of the nuclear
region. The data were divided into 3 bands, red representing 0.1 - 1.275 keV,
green 1.275 - 2.2 keV and blue 2.2 - 10 keV. There is a hard X-ray, unresolved,
core source, and extended bi-polar soft nebulosity, the latter
correlating well with the optical bi-polar continuum and emission-line
structures (e.g. Fig. 4) and being approximately aligned with the radio jet.
The bi-polar X-ray morphology may be enhanced by absorption by the dust lane
crossing the nucleus, suppressing the very soft X-ray emission.

We obtained quasi-simultaneous Chandra - RXTE observations in order to define
the broad-band spectrum. The main spectral results are: 1) the compact
nucleus is detected to 100 keV and is well described by a heavily absorbed
power law spectrum with $\Gamma_{\rm h}$ = 1.52$\pm$0.12 and equivalent hydrogen
column N$_{\rm H}$ (nuc) = 2.0$^{+0.1}_{-0.2}$ $\times$ 10$^{23}$ cm$^{-2}$.
This
column is compatible with the dust obscuration to the near infrared source
for a normal gas to dust ratio; 2) the soft ($<$ 2 keV) emission from the 
nuclear region may be described by a power law spectrum with a similar index
($\Gamma_{\rm l}$ $\simeq$ $\Gamma_{\rm h}$). Narrow emission lines from
highly ionized neon and silicon are observed in the soft X-ray spectrum of the 
NW and SE regions; 3) A ``neutral'' Fe K$\alpha$ line is detected in the nucleus
and its vicinity (r $<$ 2 kpc). The equivalent width (EW) of this line is in 
good agreement with theoretical predictions for the EW versus N$_{\rm H}$(nuc)
relationship in various geometries. An Fe K edge is also seen (Fig. 5).
Comparing this edge with N$_{\rm H}$(nuc), the nuclear iron abundance is found
to be essentially solar.

Extrapolation of the best power law model of the soft (0.5 -- 2 keV),
extended, circumnuclear emission to higher energies gives a 2 -- 10 keV
luminosity of $\simeq$ 3.7 $\times$ 10$^{42}$ erg s$^{-1}$, which is 1\% of 
that of the
unabsorbed nuclear luminosity in the same band. As noted above, 
the photon index
($\Gamma_{\rm l}$) of this soft emission agrees with that ($\Gamma_{\rm h}$) 
of the
directly viewed hard X-ray emission, and so the soft emission is consistent
with being electron-scattered X-rays from the nucleus. The scattering region
must be ionized to ionization parameter $\xi$ $\ge$ 1 in order to be
sufficiently transparent to soft X-rays. The column density of the scattering
electrons is inferred to be much lower than that required to generate the
extended
polarized optical light by electron scattering, suggesting the optical light
is, in fact, scattered by dust. The excellent spatial correlation between the
extended soft X-ray emission and the ionized gas observed optically with HST
(Fig. 4) is consistent with the soft X-ray emitting gas being photoionized.

\section{The hot spots}

Radio hot spots A, B, D and E are detected in the Chandra image (note that
the originally defined `C' is no longer considered a hot spot). Their
location and 
morphology are essentially identical to those of the corresponding radio hot
spots (e.g. Fig. 6). X-ray spectra have been obtained for the two brighter
hot spots (A and D). Both are well described by a power law with photon index
$\Gamma$ = 1.8 $\pm$ 0.2 absorbed by the Galactic column in the direction of
Cygnus A. Thermal X-ray models require too high gas densities and may be ruled
out. The images and spectra (e.g. Fig. 7) strongly support synchrotron
self-Compton models
of the X-ray emission, as proposed by Harris, Carilli \& Perley (1994) on the 
basis
of ROSAT imaging observations. Such models indicate that the magnetic field
in each of the brighter hot spots is 1.5 $\times$ 10$^{-4}$ gauss, with an
uncertainty of a few tens of percent. This value is close to the equipartition
field strengths assuming no protons are present. 

It is notable that the magnetic field
cannot be less than 1.5 $\times$ 10$^{-4}$ gauss since the SSC radiation would
then exceed the observed X-radiation. 
The alternative is that B $>$ 1.5 $\times$ 10$^{-4}$ gauss in which case the
predicted SSC emission would be too weak to account for the observed X-ray
emission. The X-rays would then have to be synchrotron radiation. However,
we feel that the synchrotron self-Compton model is by far the more likely. 
Synchrotron self-Compton emission from hot spots in radio galaxies with
magnetic fields close to equipartition has now been detected in 
3C 123 (Hardcastle, Birkinshaw \& Worrall 2001), 3C 295 (Harris et al.
2000) and Cygnus A. Thus the accumulating evidence is that equipartition
between cosmic rays and magnetic fields is common in radio hot spots.

\section{The cluster of galaxies}

As discussed above, the dominant gaseous structure is a roughly
``football shaped'' (American useage) feature with semi-major axis $\simeq
1.\!^{\prime}1$ ($\simeq 100$~kpc), which is presumably prolate
spheroidal in three dimensions. This structure apparently
represents intracluster gas which has been swept up and compressed by
a cavity inflated in this gas by relativistic material which has
passed through the ends of the radio jets. The X-ray emitting gas
shows this prolate spheroidal morphology to $\simeq 1.\!^{\prime}2$
(110~kpc) from the radio galaxy, but is spherical on larger scales.
The X-ray emission from the intracluster gas extends to at least
$8^{\prime}$ ($\simeq 720$~kpc) from the radio galaxy, and a second,
extended source of X-ray emission (probably associated with a second
cluster of galaxies) is seen some $12^{\prime}$ ($\simeq 1$~Mpc) to
the NW of Cygnus~A.
The X-ray spectrum of the integrated intracluster
gas imaged on the S3 chip (dimensions $8^{\prime} \times 8^{\prime} =
720 \times 720$~kpc), excluding the contribution from the radio galaxy
and other compact sources of X-ray emission, indicates a gas temperature,
metallicity, and unabsorbed 2--10~keV rest-frame luminosity of
$7.7$~keV, $0.34$ times solar, and $3.5 \times 10^{44}$ erg s$^{-1}$,
respectively.

The observed spectral variations in the intracluster gas are illustrated in
Fig. 8, which is a color map of the softness ratio
(1--2~keV/2--8~keV). The color image shows that emission inside
the central $\simeq 1^{\prime}$ is softer than that further out.  The
hardest emission is located at the position of the nucleus, which is
heavily absorbed (Arnaud et al. 1987; Ueno et al. 1994; Sambruna et
al. 1999; Paper II).

We have deprojected the emission from the ICM in order to measure the radial
dependence of temperature, density, thermal pressure and abundance. The
cluster was divided into 12 annuli, each centered on the radio galaxy. 
These annuli define shells in 3 dimensions. Each shell
was modeled as two mekal plasmas. First, a uniform brightness mekal
model (representing possible foreground Galactic emission) 
was added to a mekal model
for the outer shell and the result was compared with the
observed spectrum of the outer annulus. Both emission
components were assumed to be absorbed by the same column and the
model parameters for absorption and Galactic emission were
subsequently fixed at their respective best-fit values.
In this way, we derived the temperature,
abundance, and emissivity for the outer shell of intracluster gas.
Second, the spectra from the outer and adjacent shells were modeled in
the same way. The
temperature, abundance, and emissivity for the outer shell were fixed at the
values already obtained, allowing us to derive the deprojected
temperature, abundance, and emissivity for the shell adjacent to it. 
This procedure
was repeated for the remaining shells. In Fig. 9, we show the resulting 
radial dependences of gas temperature, density, metal abundance and thermal
gas pressure. It is notable that the cooling time is less than the Hubble time
for radii $\le$ 200 kpc.

Under the assumptions of hydrostatic equilibrium and spherical symmetry,
we have calculated the radial dependence of enclosed mass M($<$ r) for radii
80 - 500 kpc (Fig. 10). From this analysis, 
we find the
total mass of the cluster, within 500~kpc, is $2.0 \times 10^{14} \,
M_{\odot}$ or $2.8 \times 10^{14} \, M_{\odot}$, depending on the precise
temperature profile adopted (see Paper III).
This
compares favorably with the cluster mass of $10^{14} \, M_{\odot}$,
derived from the Einstein HRI data (Arnaud et al. 1984).
Within
500~kpc, the total mass of the intracluster gas is $1.1 \times
10^{13} \, M_{\odot}$.

\section{The Cavity and ``Belts''}

Fig. 11 is an image of the
region of the cavity. Regions for which spectra have been extracted
are indicated by the polygons. It is very interesting that all the regions in
the limb-brightened edge of the cavity are hotter than the innermost shell of
the cluster (4.9$\pm$0.6 keV, see Fig. 9, top), while all regions in the
transverse ``belts'' are cooler. While these differences in temperature are 
modest - up to 2.1 keV above and 1.1 keV below that of the innermost
cluster shell - we believe the systematic differences between the innermost 
cluster
shell, the limb-brightened regions and the ``belts'' are real. The data thus
suggest that the ICM just outside the cavity is being heated by its expansion,
as expected (e.g. Begelman \& Cioffi 1989). In contrast, the ``belts'' appear
to represent cooling gas in disk-like structures around, or within, the cavity.
The equatorial arrangement of the disks with respect to
the cavity may be interpreted in two
ways:

\noindent
(a) It is natural to speculate that the Cygnus A galaxy is accreting these gas
disks; the gas continuously cools, falls in and ultimately
fuels the accretion disk
around the black hole. In this case, the radio axis of Cyg A is
determined by the angular momentum of the large scale ICM. 

\noindent
(b) Alternatively,
the radio axis might be determined in some other way, such as the spin of the
black hole, either primordial or after accretion of gas or another black hole 
in a galaxy merger (see Stockton et al. 1994 for evidence that Cyg A has 
undergone a merger). When switched on, the radio jets power the radio lobes
and inflate the cavity, but in view of the concentration of the lobes near
the hot spots (at least at cm wavelengths), the ``robustness'' of the cavity in
the
equatorial regions may be lower than near the hot spots, allowing ICM to
fall into the cavity through
Kelvin-Helmholtz or Rayleigh-Taylor instabilities (which will mix the heavy
ICM with the light relativistic gas given the fact that gravity points
inwards). Alternatively,
even if the cavity pressure is uniform (as expected if only a relativistic
gas is present, so the cavity has a short sound crossing time), the expanding 
equatorial region may be the most unstable.

This classic ``chicken and egg'' problem is arguably the most important 
outstanding issue.

It is also of interest to determine whether there is any X-ray emitting material
{\it inside} the cavity (in addition to the obvious ``diffuse'' jet-like
features and the hot spots). As noted above, such emission  could result, for
example from ICM gas that has entered the cavity through instabilities at
the cavity - ICM contact discontinuity (e.g. Reynolds, Heinz \& Begelman
2002). To investigate this matter observationally,
it is necessary to subtract from the image the foreground and
background contributions of the cluster gas. Using our cluster model
(Section 4, see also Paper III), we assume that the axis of the prolate
structure is in the plane of the sky, the emission is cylindrically symmetric
about this axis and that the cavity is devoid of X-ray emitting gas. The 
resulting model projection
of a full band (0.75 - 8 keV) image of the cluster is shown in
Fig. 12, top. This image may then be subtracted from the observed image, the
result being shown in Fig. 12, bottom. The latter image suggests that there is,
indeed, diffuse gas within the cavity. Modelled as a thermal gas, its 
temperature
appears higher than the inner regions of the cluster
(which dominate the foreground and background emission). These 
findings should be
regarded as preliminary and the work as in progress. The main concern is that
the difference map is sensitive to systematic errors in the assumed cluster
model and geometry.

\begin{acknowledgements} 

We thank Patrick Shopbell for collaboration in the early part of this work.
This project was supported by NASA through grants NAG 81027 and NAG 81755.

\end{acknowledgements}

\clearpage

\figcaption[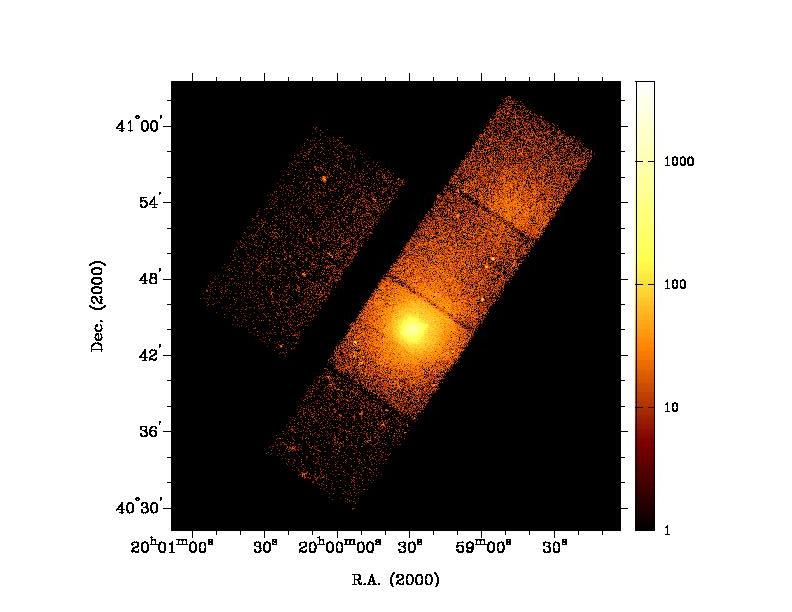]{An image of the whole \emph{Chandra} field in the
0.75--8~keV band. The image has been binned to a pixel
size of
$3.\!^{\prime\prime}94 \times 3.\!^{\prime\prime}94$.
The vertical bar indicates the conversion from color to cts
pixel$^{-1}$.\label{fig1}}

\figcaption[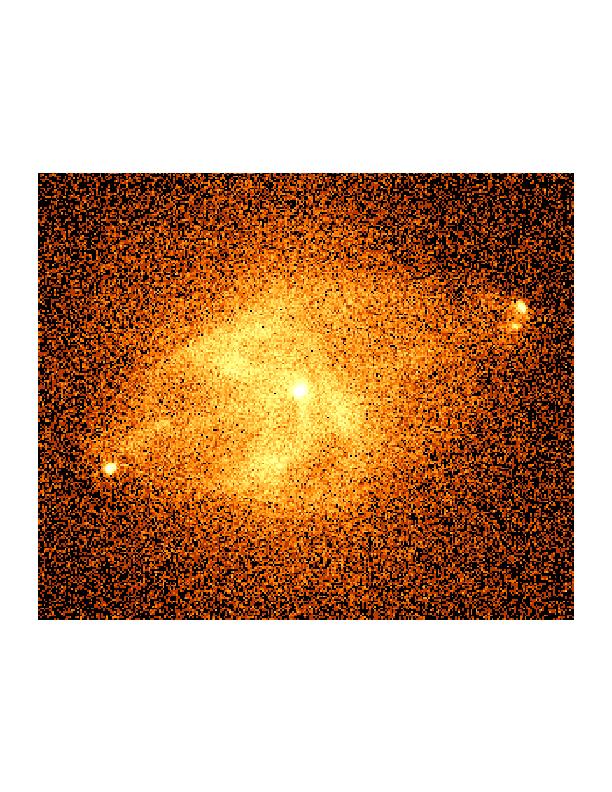]{The Chandra image on the scale of the radio source. The width
of the image is 2\farcm5. \label{fig2}}

\figcaption[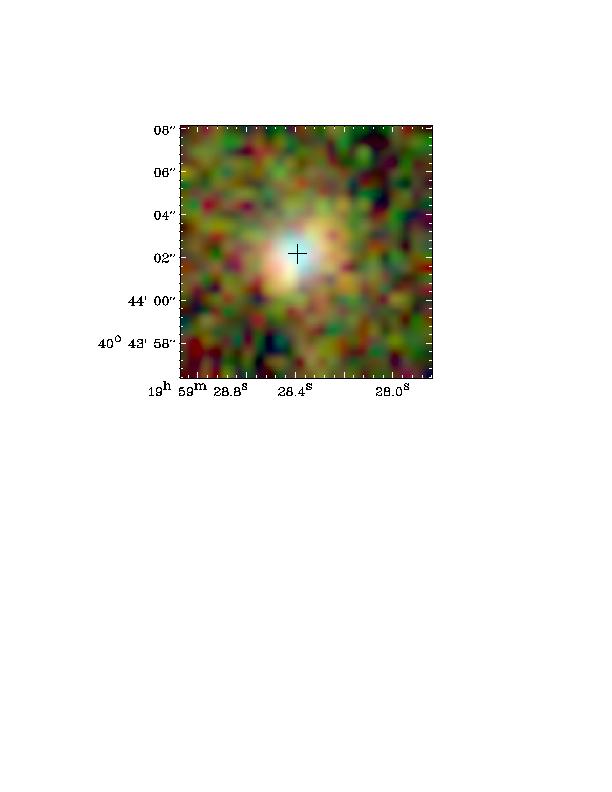]{Color representation
of the Chandra X-ray image of the nucleus of
Cyg A. The data were divided into three bands, red representing 0.1 - 1.275
keV, green 1.275 - 2.2 keV and blue 2.2 - 10 keV. The nucleus coincides with a
blue (hard) unresolved source and there is extended bi-polar soft emission.
\label{fig3}}

\figcaption[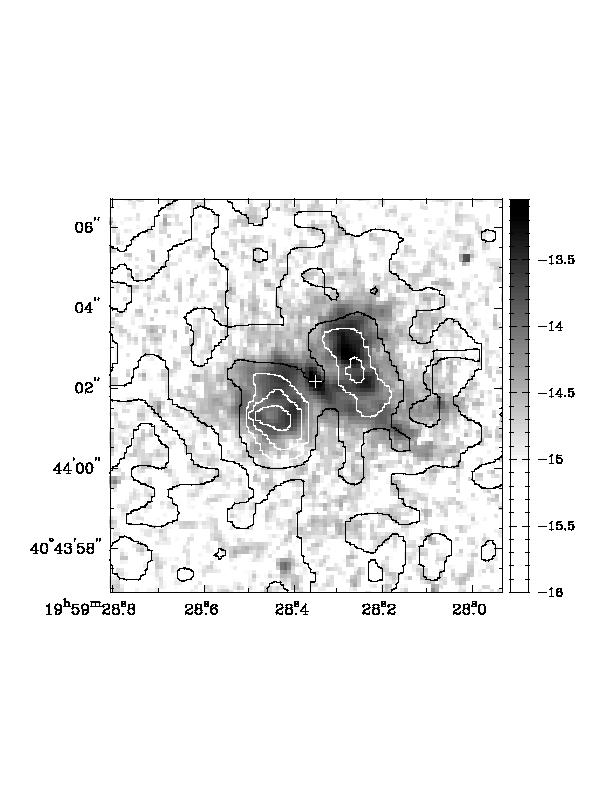]{A superposition of events in the range 0.25 to 1.00 keV (solid
contours) on an HST image taken through a Linear Ramp Filter at redshifted
[OIII] $\lambda$5007 (grey scale). \label{fig4}}

\figcaption[awilson-C2_fig5.ps]{Confidence contours (plotted at 68\%, 90\%
  and 99\%) of the rest-frame energy of the iron emission line and
  its absorption edge. The aperture is a circle
      of diameter $2\farcs5$. The line energy is consistent with that of the Fe
K$\alpha$ fluorescence line of neutral iron. The absorption edge is
consistent with that of iron in the range Fe I -- Fe X. The
panel in the upper right shows the ratio of the Chandra data to a power
law as a function of rest-frame energy. To produce this panel, the data
were
modeled by a power law from 4 to 9 keV. \label{fig5}}

\figcaption[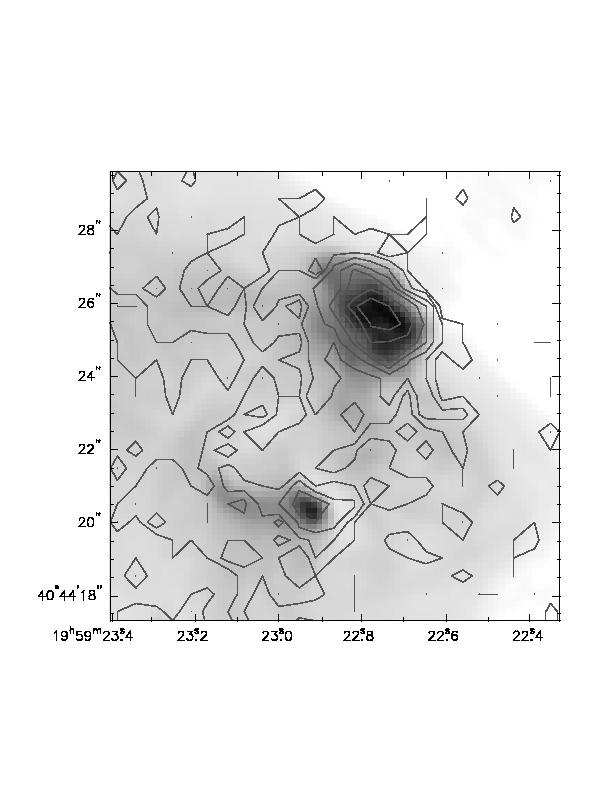]{X-ray emission 
(contours) superposed on a 6 cm VLA radio map
(grey scale)
of the region of the western hot spots (A, the brighter, and B, $\simeq$
6$^{\prime\prime}$ SE of A). Contours are plotted at 2, 4, 8, 12, 16, 24 and 32
counts per pixel (0\farcs5 $\times$ 0\farcs5). The grey scale is proportional
to the square root of the radio brightness. \label{fig6}}

\figcaption[awilson-C2_fig7.ps]
{Spectrum of hot spot A.
The points show the radio fluxes and the line through
them the model of the synchrotron radiation. The ``bow tie'' is the Chandra
measured boundary of the X-ray spectrum (these error lines are 90\%
confidence after freezing N$_{\rm H}$ at its best fit value, which coincides
with the Galactic column). The solid line is the predicted SSC spectrum
for $\gamma_{\rm min}$ = 1 and the dashed line for
$\gamma_{\rm min}$ = 100. \label{fig7}}

\figcaption[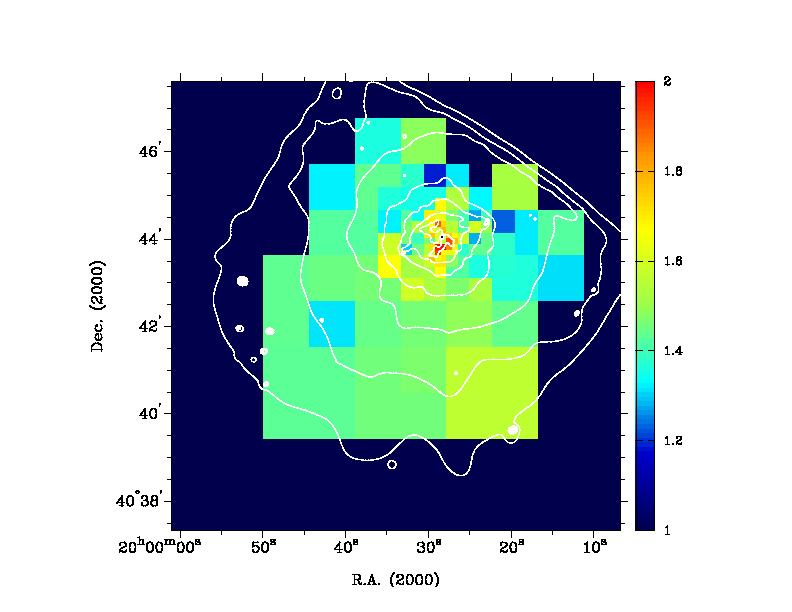]
{A color representation of the softness ratio
(i.e., 1--2~keV/2--8~keV) in the region covered by the S3 chip
superposed on X-ray contours of the background-subtracted image in the
0.75--8~keV band. The X-ray image
has been adaptively smoothed with a 2-d Gaussian profile of varying
width.  The vertical bar indicates the relation between color and
softness ratio, and the pixels were rebinned so that the fractional
error in the ratio did not exceed 0.1.  The contours indicate $2.5^{n}
\times 10^{-2}$ cts pixel$^{-1}$, where $n = 1$, 2, 3, 4, 5, 6, 7, and
8. \label{fig8}}

\figcaption[awilson-C2_fig9top.ps]
{The deprojected properties of the intracluster
gas in individual shells surrounding the radio
galaxy.  For each ellipsoidal shell, the radius is the average of the
semi-major and semi-minor axis. Top:
Gas temperature (crosses) and electron density (solid line).
Bottom: Metal Abundance (crosses) and thermal gas pressure (solid line).
\label{fig9}}

\figcaption[awilson-C2_fig10.ps]
{The integrated mass profiles M($<$ r) of the
cluster (solid and dashed lines), assuming hydrostatic equilibrium and
spherical symmetry, and the intracluster gas (dotted line) for radii
between 80 and 500~kpc.  The solid and dashed lines are for different
assumed temperature profiles (see Paper III).\label{fig10}}

\figcaption[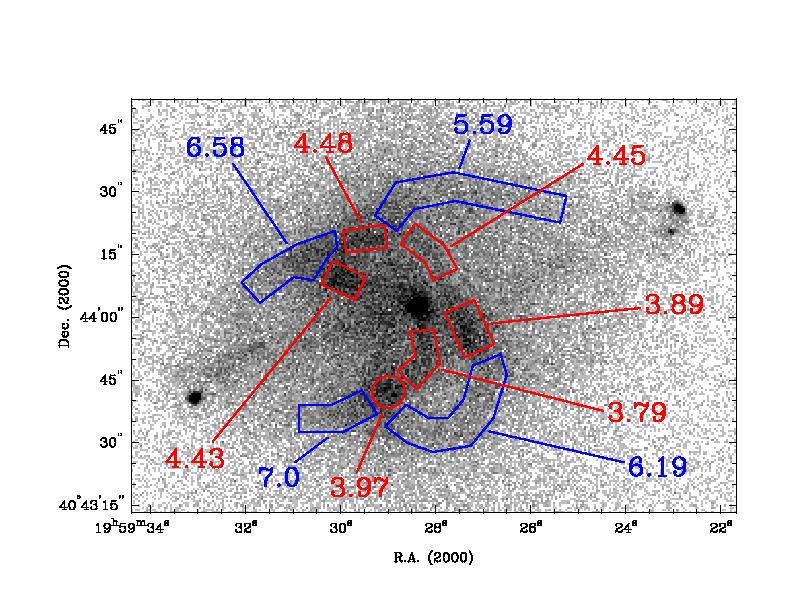]
{An unsmoothed image of the central region of the
Cygnus~A field in the 0.75--8~keV band.  The shade is proportional to
the square root of the intensity. The shading ranges from $0$ (white)
cts pixel$^{-1}$ to $15$ (black) cts pixel$^{-1}$.
The areas indicated by
solid lines
mark regions from which spectra of the X-ray emission were
extracted and modelled with mekals; the numbers indicate the resulting
gas temperature
in keV. Temperatures and regions marked in blue are hotter than the temperature
of the innermost cluster shell (4.9 keV), while those marked in red are cooler.
It is
notable that the hotter regions are associated with the limb-brightened edges
of the cavity, while the cooler regions are in the ``belts'' which encircle the
cavity approximately along its minor axis.\label{fig11}}

\figcaption[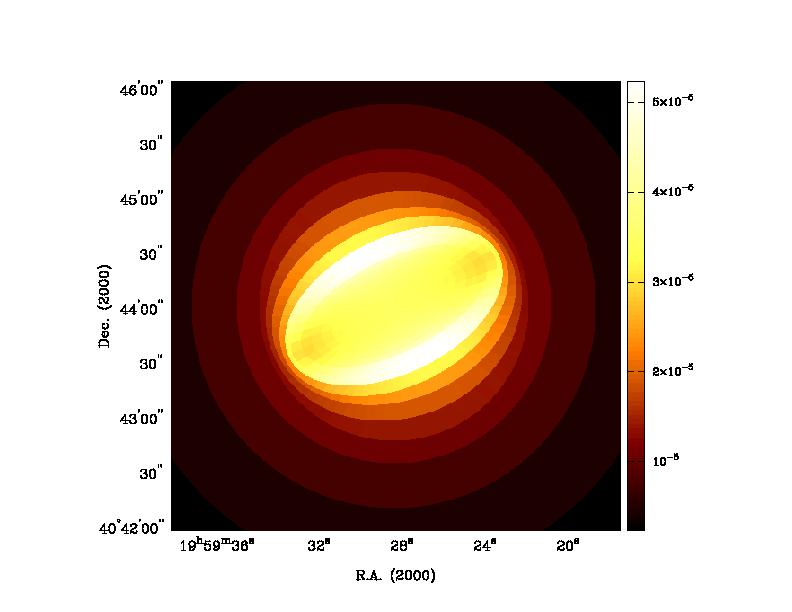]
{Top: A simulation of the observed X-ray emission in the 0.75 - 8
keV band from the model of the cluster. The cavity was assumed to radiate no
X-ray emission. Bottom: The difference between the observed image and the model
given above. Some of the black regions are negative and result from the
idealised modelling (as an ellipse) of the shape of the cavity. The color
scale is in cts s$^{-1}$ pixel$^{-1}$.\label{fig12}}


\begin{thebibliography}{}

\bibitem[\protect\astroncite{}{}]{}
Arnaud, K. A., Fabian, A. C., Eales, S. A., Jones, C. \& Forman, W. 1984,
MNRAS, 211, 981

\bibitem[\protect\astroncite{}{}]{}
Arnaud, K. A., Johnstone, R. M.,
Fabian, A. C., Crawford, C. S., Nulsen, P. E. J., Shafer, R. A., \&
Mushotzky, R. F. 1987,, MNRAS, 227, 241

\bibitem[\protect\astroncite{}{}]{}
Begelman, M. C. \& Cioffi, D. F. 1989, ApJ, 345, L21

\bibitem[\protect\astroncite{}{}]{}
Carilli, C. L. \& Barthel, P. D. 1996, Astron. Astrophys. Rev, 7, 1

\bibitem[\protect\astroncite{}{}]{}
Hardcastle, M. J., Birkinshaw, M. \& Worrall, D. M. 2001, MNRAS, 323, L17

\bibitem[\protect\astroncite{}{}]{}
Harris, D. E., Carilli, C. L. \& Perley, R. A. 1994, Nature, 367,
713

\bibitem[\protect\astroncite{}{}]{}
Harris, D. E. et al. 2000, ApJ, 530, L81

\bibitem[\protect\astroncite{}{}]{}
Reynolds, C. S. \& Fabian, A. C. 1996, MNRAS, 278, 479

\bibitem[\protect\astroncite{}{}]{}
Reynolds, C. S., Heinz, S. \& Begelman, M. C. 2002, MNRAS (in press) 
(astro-ph/0201271) 

\bibitem[\protect\astroncite{}{}]{}
Sambruna, R. M., Eracleous, M.,
\& Mushotzky, R. F. 1999, ApJ, 526, 60

\bibitem[\protect\astroncite{}{}]{}
Scheuer, P. A. G. 1974, MNRAS, 166, 513

\bibitem[\protect\astroncite{}{}]{}
Smith, D. A., Wilson, A. S., Arnaud, K. A., Terashima, Y. \& Young, A. J. 2002,
ApJ, 565, 195 (Paper III)

\bibitem[\protect\astroncite{}{}]{}
Stockton, A., Ridgway, S. E. \& Lilly, S. J. 1994, AJ, 108, 414

\bibitem[\protect\astroncite{}{}]{}
Ueno, S., Koyama, K., Nishida, M.,
Yamauchi, S., Ward, M. J. 1994, ApJ, 431, L1

\bibitem[\protect\astroncite{}{}]{}
Wilson, A. S., Young, A. J. \& Shopbell, P. L. 2000, ApJ (Letts), 544, L27 
(Paper I)

\bibitem[\protect\astroncite{}{}]{}
Young, A. J., Wilson, A. S., Terashima, Y., Arnaud, K. A. \& Smith, D. A.
2002, ApJ, 564, 176 (Paper II)


\end{thebibliography}
\end{document}